\documentclass[prl,aps,showpacs,twocolumn,superscriptaddress]{revtex4}
\usepackage{epsfig}
\usepackage{graphicx}
\usepackage{bm}

\begin{document}

\title{Proposal for Inverting the Quantum Cloning of Photons}

\author{Sadegh Raeisi}
\affiliation{Institute for Quantum Information Science, University of Calgary, Alberta T2N 1N4, Canada}
\affiliation{Institute for Quantum Computing, University of Waterloo, Ontario N2L 3G1, Canada}
\author{Wolfgang Tittel}
\affiliation{Institute for Quantum Information Science, University of Calgary, Alberta T2N 1N4, Canada}
\author{Christoph Simon}
\affiliation{Institute for Quantum Information Science, University of Calgary, Alberta T2N 1N4, Canada}

\begin{abstract}
We propose an experiment where a photon is first cloned by stimulated parametric down-conversion, making many (imperfect) copies, and then the cloning transformation is inverted, regenerating the original photon while destroying the copies. Focusing on the case where the initial photon is entangled with another photon, we study the conditions under which entanglement can be proven in the final state. The proposed experiment would provide a clear demonstration that quantum information is preserved in quantum cloning. It would furthermore allow a definitive experimental proof for micro-macro entanglement in the intermediate multi-photon state, which is still an outstanding challenge. Finally it might provide a quantum detection technique for small differences in transmission (e.g. in biological samples), whose sensitivity scales better with the number of photons than a classical transmission measurement.
\end{abstract}
\pacs{03.65.Ta, 03.65.Ud, 03.67.Mn, 42.50.Xa}


\maketitle

The no-cloning theorem \cite{nocloning} states that it is impossible to build a quantum copying machine that would perfectly copy arbitrary quantum states. This is a direct consequence of the linearity of the time evolution in quantum physics. It is also essential in order to rule out the possibility of superluminal communication using quantum entanglement \cite{superluminal}. However, approximate quantum cloning is possible \cite{buzekhillery} and has been studied extensively. Different types of quantum cloners have been introduced, including universal cloners \cite{buzekhillery,universal}, which clone all input states equally well, and phase-covariant cloners \cite{phasecovariant}, which produce equally good copies for all input states that lie on the equator of the Bloch sphere.

In all cases, the fidelity of the clones has to satisfy certain bounds, whose exact form depends on the type of cloner considered. One way of understanding these bounds is to realize that the clones cannot contain more information about the initial state than the initial state itself \cite{bruss}. One may then wonder if the clones contain exactly the same amount of information as the initial state (or less). In the case of optimal phase-covariant cloning it is easy to see that the answer is yes because the cloning transformation can be realized in a unitary fashion. It is thus in principle possible to invert the cloning transformation and recover the initial state. The answer can be shown to be yes for universal cloning as well, but one has to use more sophisticated arguments based on state estimation \cite{bruss}, as the cloner uses auxiliary systems and is thus non-unitary if only inputs and clones are considered.

Implementations of quantum cloning have been studied extensively over the last decade \cite{implementations}. A particularly simple and intuitive way of realizing quantum cloners in the context of quantum optics (where the inputs are photons) is by using stimulated emission; in this case the bounds on the fidelity of the clones can be seen as being due to the unavoidable presence of spontaneous emission \cite{stimulated}. Both universal \cite{universalexp} and phase-covariant \cite{PCexp} quantum cloners have been realized based on stimulated parametric down-conversion. However, inverting these cloning transformations has not been considered so far. The feasibility of this inversion is the topic of the present paper. Focusing on the case of phase-covariant cloning, we take into account the most important experimental imperfection, namely photon loss.

Experimentally inverting quantum cloning would be a striking demonstration of the information preservation in the cloning process. We will in particular focus on the case where the initial input photon is entangled with another photon. The preservation of this entanglement after the cloning process and its inversion is a good criterion for verifying if the original photon is indeed regenerated with its quantum character intact. However, there is another reason why this scenario is particularly interesting. There is a recent experiment \cite{DeMmicromacro} where one photon from an entangled pair was phase-covariantly cloned, and where the number of clones produced was up to tens of thousands. The authors of \cite{DeMmicromacro} claimed to have demonstrated micro-macro entanglement between one original photon and a large number of photons (the clones). This claim was subsequently challenged \cite{sekatskiPRL}, leading to a number of detailed theoretical investigations \cite{sekatskiPRA,DeMassumptions,Fabio,Raeisi}. The conclusion of this debate is that it is not easy to prove the existence of micro-macro entanglement in this system experimentally without too many assumptions. The present approach via the inversion of the cloning transformation is one possible avenue that has so far not been considered. If there is still entanglement after cloning and inverted cloning, then there definitely had to be micro-macro entanglement after the cloning step.

\begin{figure}
\epsfig{file=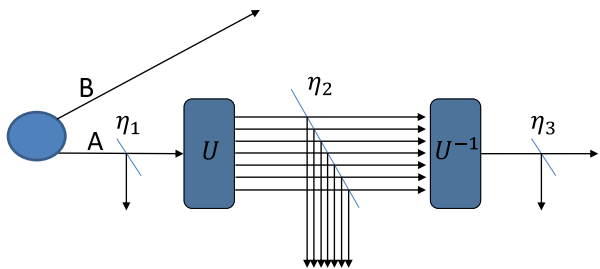,width=\columnwidth}
\caption{Setup considered in this paper. A source creates a pair of entangled photons. The photon in mode B is detected directly. The photon in mode A is cloned by the phase-covariant cloning transformation $U$, then this transformation is inverted, $U^{-1}$. The pump laser beams necessary for implementing the cloning transformations are not shown. Losses before, in between and after the cloning transformations are taken into account through the transmission factors $\eta_1$, $\eta_2$ and $\eta_3$. We are interested in the regime where the final state in mode A is again at the single-photon level. We study whether the final state of modes A and B can be shown to be entangled using the witness $W$ of Eq. (4), which is based on polarization-sensitive photon counting in both modes. The presence of (strong) entanglement between A and B in the final state can be interpreted as showing that the quantum information present in the original photon in mode A is regenerated in the final single-photon level state of mode A. Furthermore, any entanglement that is detected between A and B in the final state implies that the multi-photon state created by the first cloning transformation in A was entangled with the single photon in mode B (micro-macro entanglement), since entanglement cannot be created locally. }
\end{figure}

We now describe the system that we are considering in more detail, see also Figure 1 for the setup and Refs. \cite{sekatskiPRL,Raeisi} for the notation used. The initial photon pair is in a polarization singlet state
\begin{equation}
|\psi_-\rangle=\frac{1}{\sqrt{2}}(a_h^{\dagger} b_v^{\dagger}-a_v^{\dagger} b_h^{\dagger})|\Omega\rangle,
\label{state}
\end{equation}
 where $h$ and $v$ denote horizontal and vertical polarization and $|\Omega\rangle$ is the vacuum state for all modes. The photon in spatial mode $B$ is detected directly in a polarization-sensitive way and serves as a herald. The photon in spatial mode $A$ is subjected to the unitary phase-covariant cloning transformation $U=e^{-iHt}$, where the Hamiltonian is given by
\begin{equation}
H=i\chi a^{\dagger}_h a^{\dagger}_v + h.c.,
\end{equation}
corresponding to type-II collinear parametric down-conversion \cite{PCexp,DeMmicromacro}. The coupling constant $\chi$ includes the non-linear coefficient of the crystal and the amplitude of the pump laser. The spatio-temporal mode $a$ in the Hamiltonian has to be indistinguishable from that of the input photon in order for stimulated emission to occur \cite{stimulated}. Identifying $h$ and $v$ with the north and south poles of the Bloch sphere and introducing equatorial modes $a_{\phi}=\frac{1}{\sqrt{2}}(e^{i\frac{\phi}{2}} a_h+e^{-i\frac{\phi}{2}} a_v)$ and $a_{\phi \perp}=a_{\phi+\pi}$, one has
\begin{equation}
H=\frac{i\chi}{2}(a^{\dagger 2}_{\phi}+a^{\dagger 2}_{\phi \perp})+h.c. \label{squeezers}
\end{equation}
The Hamiltonian thus corresponds to a sum of two squeezers for any two orthogonal equatorial modes. As a consequence, $U$ factorizes into two independent unitaries, one for $a_{\phi}$ and one for $a_{\phi \perp}$. Note that since the choice of $\phi$ is arbitrary, we will use the notation $a$ and $a_{\perp}$ for the equatorial modes for simplicity.

The inverted cloning transformation $U^{-1}$ can be implemented by changing the sign of $\chi$. Physically this can be done by changing the phase of the pump beam for the down-conversion process. Note that if $U^{-1}$ acted on a single-photon input state, it would create a large number of clones in full analogy with $U$, with changes only in certain phase factors that depend on the sign of $\chi$. However, acting after $U$ it has the effect of converting a multi-photon state back to the single-photon level.

We want to study the entanglement in the final state, i.e. after cloning and inverted cloning. In the absence of imperfections, inverting $U$ trivially leads one back to the initial state. The situation is more interesting when realistic imperfections are taken into account. We consider photon loss before $U$, between $U$ and $U^{-1}$, and after $U^{-1}$, characterized by transmission coefficients $\eta_1$, $\eta_2$ and $\eta_3$, see Figure 1. We will see that all three types of loss affect the entanglement, but it is clear that the loss between $U$ and $U^{-1}$ plays a special role, because it prevents the cancelation of $U$ and $U^{-1}$. We are interested in the regime where the cancelation is still close to perfect, such that the final state in mode $A$ is at the single-photon level, but may nevertheless contain significant vacuum and few-photon components. We therefore need an entanglement witness that can deal with these components. We use the entanglement criterion of Ref. \cite{sekatskiPRL}, which is based on that of Ref. \cite{entlaser}. The state is proven to be entangled if
\begin{equation}
W=|\langle \vec{J}^A \cdot \vec{\sigma}^B \rangle|-\langle N^A \rangle >0,
\end{equation}
where $\vec{J}^A$ is the vector of Stokes operators for mode $A$, i.e. $J_z^A=a^{\dagger}_h a_h - a^{\dagger}_v a_v$, $J_x^A=a^{\dagger}_0 a_0-a^{\dagger}_{0\perp} a_{0\perp}$, $J_y^A=a^{\dagger}_{\frac{\pi}{2}} a_{\frac{\pi}{2}}-a^{\dagger}_{\frac{\pi}{2}\perp} a_{\frac{\pi}{2}\perp}$, where $a_0$, $a_{0\perp}$, $a_{\frac{\pi}{2}}$ and $a_{\frac{\pi}{2}\perp}$ are equatorial modes as introduced above; $\vec{\sigma}^B$ is the corresponding vector of Stokes operators for mode $B$, but restricted to the single-photon subspace; and $N^A=a^{\dagger}_h a_h + a^{\dagger}_v a_v$ is the total number of photons in $A$.

The multi-photon states generated by the cloning transformation $U$ are quite complex \cite{DeMmicromacro,sekatskiPRL,sekatskiPRA}. It is therefore much more convenient to work in the Heisenberg picture, i.e. to transform the operators and evaluate their expectation values for the initial two-photon singlet state. Applying the appropriate sequence of beam splitter and squeezer operations corresponding to Figure 1, we find the overall transformation for any equatorial mode,
\begin{eqnarray}
a'=\sqrt{\eta_1\eta_2\eta_3} \, a + \sqrt{(1-\eta_1)\eta_2\eta_3} \, c_1 +\nonumber\\
\sqrt{(1-\eta_2)\eta_3}(\cosh(g) \, c_2 - \sinh(g) \, c_2^{\dagger}) + \sqrt{1-\eta_3} \, c_3,
\label{trf}
\end{eqnarray}
where $a'$ is the final mode, $a$ is the initial mode, $c_1, c_2, c_3$ are the vacuum modes injected into the system by the losses before, in between, and at the end (see Figure 1); $g=\chi t$ is the gain of the cloner (and of the inverted cloner). One can see that $c_1$ and $c_3$ just cause loss. In contrast the vacuum fluctuations due to $c_2$ are amplified by the second (inverting) cloner. This shows clearly that the intermediate loss is particularly important in the present context.

In order to calculate expectation values in the Heisenberg picture we also need the initial state. It is given by Eq. (\ref{state}), which corresponds to the vacuum state for all the loss modes. It is helpful to rewrite it in terms of equatorial modes $a$ and $a_{\perp}$ as $|\psi_-\rangle=\frac{1}{\sqrt{2}}(a^{\dagger} b^{\dagger}_{\perp}-a^{\dagger}_{\perp} b^{\dagger})|\Omega\rangle$, using the simplified notation that we introduced after Eq. (\ref{squeezers}).

It is not hard to show that the final mean photon number in $A$ is
\begin{eqnarray}
\langle \psi_-| N^A |\psi_- \rangle = \langle 1|a'^{\dagger} a' |1\rangle + \langle 0| a'^{\dagger} a'|0\rangle = \nonumber\\
\eta_1 \eta_2 \eta_3 + 2 (1-\eta_2) \eta_3 \sinh^2(g), \label{NA}
\end{eqnarray}
where $|0\rangle$ and $|1\rangle$ are the zero-photon and one-photon states for the initial mode $a$. We are interested in the regime where the transmission factors $\eta_i$ are all fairly close to one (with $\eta_2$ very close to one, see below), such that the $\eta_1 \eta_2 \eta_3$ term in Eq. (\ref{NA}) is of order one, and where $2 (1-\eta_2) \eta_3 \sinh^2(g)$ is small compared to $\eta_1 \eta_2 \eta_3$, such that the final state in $A$ is again at the single-photon level. That is, the regime where the effect of the amplified vacuum fluctuations discussed above is relatively small.

Turning now to the question of entanglement in the final state, using Eq. (\ref{trf}) one can show that
\begin{eqnarray}
\langle \psi_-| J_x^A \sigma_x^B |\psi_- \rangle=\langle \psi_-| J_y^A \sigma_y^B |\psi_- \rangle=\nonumber\\
-\left( \langle 1|a'^{\dagger} a' |1\rangle - \langle 0| a'^{\dagger} a'|0\rangle \right)=-\eta_1 \eta_2 \eta_3.
\end{eqnarray}
Furthermore
\begin{equation}
\langle \psi_-| J_z^A \sigma_z^B |\psi_- \rangle=-\eta_1\eta_2\eta_3.
\end{equation}
Note that $J_z^A$ commutes with $H$ and is thus not affected by the cloners at all. Putting all these pieces together, one finds
\begin{eqnarray}
W= 2 \left( \eta_1 \eta_2  - (1-\eta_2) \sinh^2(g)\right) \eta_3.
\label{W}
\end{eqnarray}
From Eq. (\ref{W}) one can see that different loss channels have quite different effects on the entanglement witness, as was to be expected following the discussion after Eq. (\ref{trf}). Loss after the second cloner affects both terms equally and is thus relatively benign. The value of the witness just
decreases proportionally to the transmission $\eta_3$. Loss before the first cloner reduces the first term, but only linearly in $\eta_1$. Note that this is in contrast to the situation for just one cloner, where loss before the cloner greatly affects the violation of the same entanglement criterion, see section VI.C of Ref. \cite{sekatskiPRA}. This difference is due to the fact that in the present situation errors due to loss before the first cloner (which can be viewed as fluctuations due to the injected vacuum mode) are amplified by the first cloner, but de-amplified by the second one. The main problem in the present situation is loss between the two cloners, characterized by $\eta_2$. Errors due to this intermediate loss  are amplified by the second cloner by a factor $\sinh^2(g)$. The size of this amplification is directly related to the mean number of photons (clones) after the first cloner, which is
\begin{equation}
N^A_c=\langle \psi_- | \tilde{a}^{\dagger} \tilde{a}|\psi_- \rangle= 2 (1+\eta_1) \sinh^2(g) + \eta_1, \label{NAc}
\end{equation}
where the subscript $c$ is for ``clones'' and $\tilde{a}=(\sqrt{\eta_1} a+ \sqrt{1-\eta_1} c_1) \cosh(g)+(\sqrt{\eta_1} a^{\dagger}+ \sqrt{1-\eta_1} c_1^{\dagger}) \sinh(g)$.

\begin{figure}
\epsfig{file=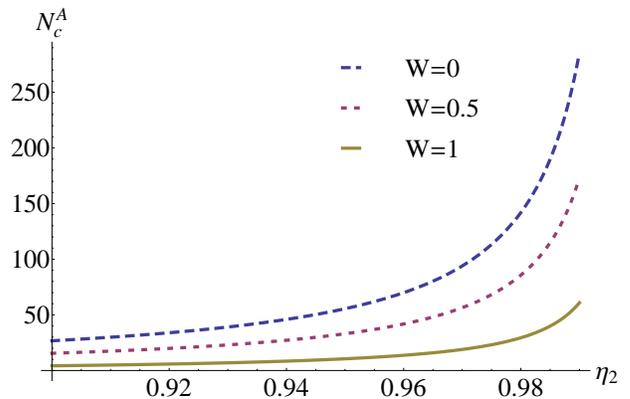,width=\columnwidth}
\caption{The number of clones after the first cloner $N_c^A$ for which the entanglement witness $W$ for the final state takes values 0, 0.5, 1 (from top to bottom), as a function of the intermediate transmission $\eta_2$, for fixed $\eta_1=\eta_3=0.8$. Points below the top curve correspond to entanglement. Note that the theoretical maximum value for the witness is 2. As a consequence of Eqs. (9-10), proving entanglement is more difficult for larger numbers of clones in the intermediate state (i.e. for higher gain), requiring values of $\eta_2$ increasingly close to 1.}
\end{figure}

Eqs. (\ref{W}) and (\ref{NAc}) give
\begin{equation}
W=\frac{\eta_1 (1+\eta_2 + 2\eta_1 \eta_2) \eta_3}{1+\eta_1}-\frac{(1-\eta_2)\eta_3}{1+\eta_1} N_c^A.
\label{W-N}
\end{equation}
The negative term in this expression for $W$ is proportional to both $(1-\eta_2)$ and $N_c^A$. This implies that in order to prove entanglement using the witness $W$ for increasing intermediate numbers of clones $N_c^A$, the transmission $\eta_2$ has to be closer and closer to one. This is illustrated in Figure 2. This probably rules out detecting entanglement for photon numbers of order $10^4$, as in the experiment of Ref. \cite{DeMmicromacro}. Supplementary assumptions thus would still have to be made in order to prove entanglement for such large photon numbers, see also the discussion in the introduction and Refs. \cite{DeMmicromacro,sekatskiPRL,sekatskiPRA,DeMassumptions,Fabio,Raeisi} .
However, the approach described here should allow to demonstrate entanglement for
 photon numbers much bigger than one without any additional assumptions, see Figure 2. Using anti-reflection coatings it should be possible to keep the losses between the two cloners, i.e. $1-\eta_2$, at the level of at most a few percent. The transmission ``before'', $\eta_1$, is equivalent to the heralding efficiency for single-photon sources based on parametric down-conversion, for which values as high as 0.83 have been reported \cite{pittman}. The transmission ``after'', $\eta_3$, is mainly limited by the detection efficiency, for which values as high as 0.95 have been achieved using transition-edge sensor detectors \cite{TES}. In Figure 2 we have conservatively assumed $\eta_1=\eta_3=0.8$. The figure shows that micro-macro entanglement involving a hundred photons or more on the macro side should be provable with the present approach, which therefore provides a new avenue for proving micro-macro entanglement without supplementary assumptions.

The high sensitivity to the transmission $\eta_2$ (in the high photon-number regime) shown here is closely related to the results of Refs. \cite{Fabio} and \cite{Raeisi}, which show that the detectable micro-macro entanglement is highly sensitive to loss (for a homodyne measurement) and coarse-graining (for photon counting measurements) respectively. Let us note that micro-macro entanglement is present even for small values of $\eta_2$ (i.e. large loss), at least for the simple model used in this paper. This follows from the results of Ref. \cite{sekatskiPRL}, where the same entanglement witness that we use here was applied directly to the micro-macro state. However, the presence of this entanglement can be proved experimentally only by very difficult measurements, which involve counting large photon numbers with
single-photon level resolution \cite{Raeisi}.

In the above discussion we have assumed that the gain values for the two cloners are exactly identical. Experimental imperfections are likely to result in a small difference, such that the first cloner has a gain $g_1$ and the second cloner has a gain $g_2=g_1+\epsilon$. One can show that in this case the value of the entanglement witness in the final state is $W'=W-\eta_1 \eta_2 \eta_3 \epsilon^2$, where $W$ is the value in the ideal case discussed above. There is no correction term linear in $\epsilon$, making the violation relatively robust, because the case of perfectly matched gains is in fact an optimal point that maximizes the final entanglement.

So far we have discussed the proposed experiment from a purely foundational point of view, as a way to demonstrate information preservation in quantum cloning and the existence of micro-macro entanglement. We now argue that it might also be interesting from a much more applied perspective. Detecting small variations in transmission across a sample is one of the most fundamental problems in optical imaging. For example, biological samples often have very low contrast, see e.g. \cite{pawley} for more details. The standard approach is to use a classical beam of light (corresponding to a coherent state in the quantum description) and measure the variation of the transmitted intensity. However, a coherent state has a Poissonian photon number distribution, which implies that it has photon number fluctuations of $\sqrt{N}$, for a mean number of photons $N$ in the beam. This means that a small change in transmission $\Delta \eta$ is only observable if $N \Delta \eta > \sqrt{N}$, or $\Delta \eta > \frac{1}{\sqrt{N}}$. For small $\Delta \eta$ one thus requires quite large $N$, which can be a problem for highly light-sensitive samples (e.g. living cells).

In contrast, consider a situation where the sample is placed between the first and second cloner in Figure 2. Eq. (\ref{W-N}) implies that $W$ varies strongly with a change in $\eta_2$,
\begin{equation}
\frac{dW}{d\eta_2}=\frac{\eta_3}{1+\eta_1} N_c^A + \frac{\eta_1 \eta_3(1+2\eta_1)}{1+\eta_1}.
\end{equation}
The first term dominates over the second one even for quite modest intermediate photon numbers $N_c^A$. This means that $\frac{dW}{d\eta_2}$ is {\it linear} in the total number of photons $N_c^A$ transmitted through the sample. For comparison with the above discussion of the coherent-state case we will refer to this number as $N \equiv N_c^A$. The smallest detectable change $(\Delta \eta_2)_{min}$ is given by $(\Delta W)_{min}/(\frac{dW}{d\eta_2})$, where $(\Delta W)_{min}$ is the smallest detectable change in $W$. Under typical experimental conditions $(\Delta W)_{min}$ will depend on the precision of the experiment and on the number of repetitions, but it will be independent of $N$. As a consequence, for large enough $N$, $(\Delta \eta_2)_{min}$ scales like
$\frac{1}{N}$, compared to $\frac{1}{\sqrt{N}}$ in the case of the coherent state.
This much more advantageous scaling is a quantum enhancement that is due to the use of entangled light. It is analogous to the Heisenberg limit (in contrast to the standard quantum limit) in interferometry \cite{giovannetti,davidovich}. This suggests that the present approach may be promising as a quantum measurement technique.

This work was supported by AITF, NSERC and General Dynamics Canada.

\end{document}